\documentclass[twocolumn,secnumarabic,amssymb, nobibnotes, showpacs, showkeys, aps, prc]{revtex4-1}
\usepackage{balance}
\usepackage{amsmath} 
\usepackage{amsfonts}
\usepackage{amssymb} 
\usepackage[active]{srcltx} 
\usepackage[dvips]{hyperref} 
\usepackage[dvipdfm]{graphicx} 
\usepackage{float}
\newcommand {\BF}[1] {\mbox{\boldmath ${#1}$}}

\begin{document}

\title{Role of pairing in the description of Giant Monopole Resonances}
\author{Paolo Avogadro and Carlos A. Bertulani}
 \affiliation{ Texas A\&M University-Commerce, Department of Physics and Astronomy, P.O. Box 3011, 75428, Commerce (Tx), USA  }

\date{\today}

\begin{abstract}

We compare the results obtained in the framework of the quasiparticle random phase approximation (QRPA) on top of a Hartree-Fock-Bogoliubov (HFB) with the most recent experiments on giant monopole resonances in Pb, Sn, Zr, Sm, Mo and Cd.
Our calculations are fully self consistent and the density dependence of pairing interactions is, for the first time in this framework, properly taken into account. 
 In the particle-hole (ph) channel we employ different Skyrme functionals (SLy5, SkM* and Skxs20) while in the particle-particle (pp) channel we make use of density dependent contact interactions.  We introduce in the pp channel the recently proposed contact interactions which take into account the neutron-proton asymmetry. We find that no single parametrization is able to reproduce with sufficient accuracy all the nuclei. Since about two-thirds of the nuclei under investigation are better explained with a soft parametrization this tends to suggest that the currently accepted picture for the incompressibility might require modifications.
\end{abstract}


\maketitle


\begin{section}{Introduction}\label{intro}
The study of the isoscalar giant monopole resonance (ISGMR) has been for a long time \cite{first} a valuable tool to investigate
 the nuclear structure and to extrapolate the behavior of nuclear matter \cite{RS}. Various approximations 
 in the energy density functional (EDF)
framework using interactions with nuclear matter incompressibility modulus $K_{NM}\approx 230\pm 20$ MeV have shown good agreement with the experimentally determined centroids, in particular for $^{208}$Pb, $^{90}$Zr, and $^{144}$Sm. 
In Ref. \cite{12Past} Niksi{\'c} \textit{et al.} showed for the first time that a relativistic mean field theory approach is also able to return good estimates of the ISGMR centroids with the same value of the incompressibility (improving older results which required values of $K_{NM}\approx$ 250 MeV). This further confirmation seemed to settle down the problem.

More recent experimental data on Sn \cite{Sn-exp-osaka} and Cd \cite{exp-cadmium2, exp-cadmium3} however, cannot be reproduced  equally well with the same functionals. The differences between the experimental and the calculated centroids are about 0.5 - 1.0~MeV; for this reason the compressibility of atomic nuclei has become once again an important topic in nuclear structure~\cite{Pieka07}. 

Many attempts have been made to reduce the gap between experiments and theory. In a work by Col{\`o} \textit{et al.} \cite{SymColo} it has been pointed out that increasing the value of the symmetry energy J at saturation density  can decrease the calculated centroids while keeping the incompressibility constant.  

The role of superfluidity on the softness of nuclei has been investigated in an early work by Civitarese \textit{et al} in \cite{ring-pairing} and more recently it has been addressed in detail by Khan in \cite{Khan09}. It is now generally accepted that pairing can help in reducing the problem \cite{colo+hfb+qrpa}, but cannot cure it completely. For this reason, a prominent role in studying giant monopole resonances is played by the harmonic approximation, taking into account the pairing correlations (QRPA). 
It is important to remark that neglecting parts of the energy density functional in the residual fields can lead to incorrect estimates of the centroids of the ISGMR \cite{shlomo-self}. 
 A completely self consistent QRPA on top of a HF+BCS and its extension quasiparticle time blocking approximation were adopted in \cite{13Past}, but also in this case a good agreement with the experiments required different Skyrme parametrizations according to the isotopic chain. The same framework (HF+BCS+QRPA) but with density dependent delta interactions (DDDI) in the pairing channel and other kinds of Skyrme interactions in the mean field have been used by Cao \textit{et al.} in \cite{colo+cadmium}.

The HFB, however, is a more general approach to pairing properties than HF+BCS (in particular for neutron rich nuclei) and it should to be preferred as a basis for the QRPA \cite{Khan09}. 
 In the case of ISGMR, it has been shown that the two approaches can give rise important differences in the determination of the centroids \cite{colo+hfb+qrpa}. 
In the same article \cite{colo+hfb+qrpa} Sn isotopes ISGMR centroids were studied in the framework of a HFB+QRPA with the use of DDDI interactions, however both of the rearrangement terms due to the density dependence of the pairing interaction and the contribution of the two-body spin-orbit interaction were neglected.

  A systematic fully self consistent HFB+QRPA calculation (solved with Arnoldi method) of ISGMR was shown in \cite{pastDoba}. In this paper the effects of a separable interaction in the pairing channel were compared to a \textit{volume} type pairing. This kind of pairing interaction, however, also fails at curing the discrepancies between the experiments and the calculations.  

A clear assessment on the validity of the Skyrme energy EDF and density dependent pairing framework requires thus the least number of approximations: fully self consistent HFB+QRPA+DDDI.

The scope of this paper is to clarify via systematic calculations which are the most suitable parametrizations able to reproduce the centroids of the ISGMR. In order to achieve this result we employ an accurate and fully self consistent HFB+QRPA method. To our knowledge, this is the first calculation of this kind that properly takes into account the rearrangements due to the density dependence of the pairing interaction in the context of the HFB+QRPA. 

Our calculations are based on the finite amplitude method (FAM) \cite{fam, qfam} and in particular on the recent matrix-FAM \cite{mFam}. This method has proven to return results as accurate (if not more) as the QRPA expanded on a canonical basis, since the latter requires special care in handling the canonical basis \cite{Terasaki1}, and additional cutoffs are required \cite{Terasaki1,colo+hfb+qrpa} in building the QRPA matrix to prevent it to become too large. These problems are completely avoided with the FAM using a HFB in the coordinate space. \\

 Section \ref{sec:detail} is devoted to the details of the calculations.
 From Sec. \ref{sec:lead} to Sec. \ref{sec:Cadmium} we show the results obtained with the popular density dependent pairing interactions 
of the $volume$, $surface$ and $mixed$ type for Pb, Sn, Zr, Sm, Mo and Cd isotopes, and we compare them with experiments.  Sec. \ref{sec:confrontoColo} explains the main differences between our results and the ones reported in \cite{colo+cadmium}.
In Sec. \ref{sec:isovector-pairing} we include the newly proposed \cite{MSH,YS} density dependent isovector interactions and we address the impact they have on the softness of superfluid nuclei. In Sec. \ref{sec:Pairing} we modify the parameters of the pp interaction to a wide range of values to test if any parametrization can be able to solve the puzzle. The last section is devoted to the conclusions.\\
\end{section}
\section{Detail of the calculations}\label{sec:detail}
In order to compare the theoretical and experimental data, we are interested in the centroid of the ISGMR. This quantity can be directly related
with the nuclear incompressibility modulus $K_A$. One can consider $K_A$ as the analogous for a nucleus of the elastic constant of a spring; for this reason we will refer to nuclei with a low value of $K_A$ as soft or to the ones with high values of $K_A$ as stiff. Because of the definition of $K_A$ this means that the higher the centroid the stiffer the nucleus \cite{blaizot80}:
\begin{equation}
  E_{ISGMR} =\hbar \sqrt{\frac{ K_A}{m\langle r^2 \rangle}},
\end{equation}
$m$ being the nucleon mass, $\langle  r^2 \rangle $ the mean square radius of the nucleus, and $E_{ISGMR}$ being the centroid energy.
It has been shown that extracting the value of the incompressibility of nuclear matter $K_{NM}$ as the volume term of a leptodermus formula from $K_A$ is strongly model dependent \cite{blaizot80, ShlomoYoung}. For this reason, the most reliable way to obtain $K_{NM}$ is to find a microscopic model able to reproduce accurately the experimental results for a nucleus or a series of nuclei; the same microscopic framework can then be used to obtain the equation of state of nuclear matter from which $K_{NM}$ is defined as \cite{blaizot80}:
\begin{equation}
 K_{NM}= 9 \rho_0 \left. \frac{\partial^2 E/A }{\partial \rho^2} \right|_{\rho_0},
\end{equation}
where $\rho_0$=0.16 fm$^{-3}$ is the saturation density and $E$ is the energy of the system.
Although the present work is devoted to finite nuclei, the incompressibility of nuclear matter $K_{NM}$ is a universal quantity (not dependent on the single nucleus like $K_A$) and we will use it to compare the results of different EDF. 

In order to extract the correct value for the centroid of the giant resonance ($E_{ISGMR}$)
one can use the ratios $m_1/m_0$, $\sqrt{m_1/m_{-1} }$ and $\sqrt{m_3/m_1}$. Since in literature the most reported estimates of the centroid are based on $m_1/m_0$, we will focus on this quantity. The moments are defined as:
\begin{equation}\label{eq:moment}
 m_k= \int_0^{\infty} E^{k} S(E)dE, 
\end{equation}
where the strength is $S(E)= \sum_j |\langle 0|F_0|j \rangle|^{2}\delta(E-E_j)$; $|0\rangle$ being the ground state and $|j\rangle$ an eigenstate with energy $E_j$ of the QRPA. 
The monopole operator is of the form:
\begin{equation}
F_0 = \sum_{i=1}^{A} r_i^2 .
\end{equation}
 In theory one should calculate the integral of Eq. (\ref{eq:moment}) from zero to infinity. %
In practice, the QRPA calculations return discrete values for the eigenstates and the integral of Eq. (\ref{eq:moment}) reduces to a finite summation.
When trying to compare the experimental results and theoretical findings, one has to take into
account that the values of the centroid can be strongly affected by the energy interval where the moments are calculated, for this reason we will report it in all our results. 

With the finite amplitude method an existing HFB code is required to build the QRPA code \cite{qfam}.
 We employ the \textit{HFBRAD} \cite{BennaDoba}, which solves the HFB equations in coordinate space in a spherical box.
We modified the \textit{HFBRAD} in order to introduce interactions which were not present in the original version.
 The maximum angular momenta
taken into account are $J_{max}$=21/2, 15/2
for neutrons and protons.
The HFB self consistent iterations are considered convergent when the relative difference between the calculated energies is smaller than $10^{-9}$ and the maximum variation of the sum of the neutron and proton pairing gap is smaller than $10^{-7}$ MeV. 
The HFB calculations are performed with \textit{quadruple} precision real numbers, while the QRPA matrix construction and diagonalization with \textit{double} precision variables.
For the mean field we use SLy5 \cite{ref-sly}, SkM* \cite{ref-skm} and Skxs20 \cite{ref-skx} Skyrme interactions (this last interaction has been created to replace the SkP, which, according to \cite{ref-err-skp} presents intrinsic stability problems and it is thus unreliable). In accordance to the fact that the SLy5 interaction has a value of $K_{NM} \approx 230$ MeV, while SkM* has $\approx$ 217 MeV and Skxs20 have $\approx$ 202 MeV, these parametrizations cover a wide range of possible incompressibilities. In Appendix \ref{App:B} we report the main properties associated to the interactions used and we compare the charge radii and binding energy with the experimental values of the nuclei under investigation.  

In the pairing channel we apply density dependent delta interactions.
 Due to the divergent character of a delta force we limit the number of states used for the calculations  (see later for further details).
The pairing interaction present in the \textit{HFBRAD} is of the form \cite{BennaDoba}:
\begin{equation}\label{eq:pairing-int}
 v({\bf r},{\bf r'}) = V_0\left[1-\eta \left( \frac{ \rho} {\rho_0} \right)^{\gamma} \right]\delta({\bf r}-{\bf r'}).
\end{equation} 
The density dependence of the interaction mimics the pairing suppression at high momenta (density). According to how the parameters $V_0, \eta$ are modified one can obtain different kinds of pairing called $volume$, $surface$ and $mixed$ pairing. The names reflect the pairing field they give rise to, which can be localized in the volume, on the surface, or on a mix of the two. The \textit{volume} interaction in particular has no explicit density dependence ($\eta=0$), and for this reason it is easier to treat it when calculating the residual fields.
 As a representative of the \textit{mixed} pairing case we choose $\eta = 0.5$.  

Among the different types of pairing, \textit{surface} pairing is expected to have a special role; in fact
the compressibility is particularly sensitive to the surface properties of a nucleus \cite{blaizot80}. 
Theoretical models show that the coupling with collective vibrations leads to 
pairing fields strongly peaked at the surface of the nucleus \cite{pastEnrico} and
 according to Ref. \cite{bertsch2009} \textit{surface} pairing can reproduce nuclear masses with better accuracy with respect to other parametrizations. 
 In order to have a pairing field peaked on the surface and vanishing in the central part of a nucleus where $\rho=\rho_0$, according to Eq. (\ref{eq:pairing-int}), $\eta$ is set equal to 1.
In these calculations $\gamma$=1 (we will test in Sec.~\ref{sec:Pairing} pairing interactions with different values of $\gamma$). In Fig. \ref{fig:Pairing-comparison} (b) we show different  kinds of pairing as a function of the parameter $\eta$ passing, with small steps, from \textit{volume} to \textit{surface} pairing.
\subsection{Rearrangement}
\begin{figure}[ht!]
 \centering
 \includegraphics[trim = 7mm 4mm 5mm 5mm, width=8cm]{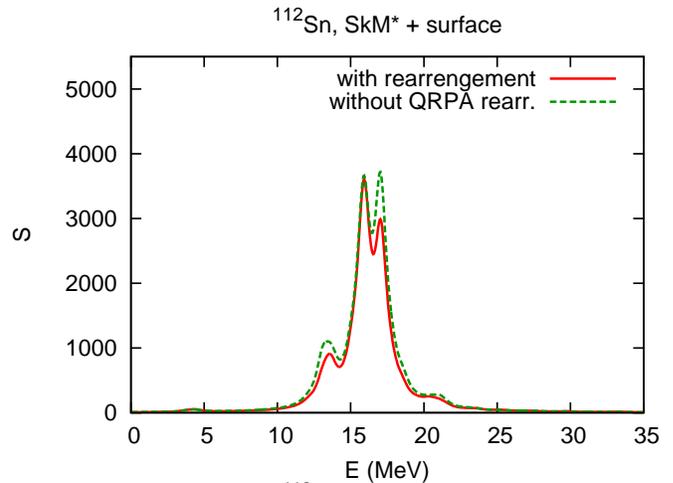}
 \caption{(Color online) $^{112}$Sn strength functions calculated with SkM*+ \textit{surface} pairing, including the rearrangement term or neglecting it only in the QRPA calculations. The strength function is in arbitrary units.}
  \label{fig:rearr}
 \end{figure}

The rearrangement term due to the density dependence of the pairing interaction is properly taken into account in this work. This term is originated by the fact that the energy in a Skyrme and DDDI approximation has the form: $ \mathcal{E}= \mathcal{E}_{kin}+ \mathcal{E}_{Skyrme}+ \mathcal{E}_{pair} + \mathcal{E}_{coul}$,
where the kinetic term, the Skyrme term, the pairing term and the Coulomb term are present. 
In the HFB approach the single particle Hamiltonian is obtained as a functional derivative of the energy with respect to the density; the part of the single particle Hamiltonian coming from the $\mathcal{E}_{pair}$ is usually referred to as rearrangement term:
\begin{equation}
 h=                                           \frac{\delta \mathcal {E}_{kin}}{\delta \rho} 
                                            + \frac{\delta \mathcal {E}_{skyrme}}{\delta \rho}
                                            + \frac{\delta \mathcal {E}_{pair}}{\delta \rho}
                                            + \frac{\delta \mathcal {E}_{coul}}{\delta \rho}.
\end{equation}
Similarly, the residual fields giving rise to the QRPA matrix are functions of the second derivatives of the densities and the rearrangement term is:
\begin{equation}
 \frac{\delta h_{rearr}}{\delta \rho} = \frac{ \delta } {\delta \rho } \left(  \frac{\delta \mathcal {E}_{pair} }{\delta \rho} \right). 
\end{equation}
If there is no explicit density dependence of the pairing term, $\mathcal{E}_{pair} \neq \mathcal{E}_{pair}[\rho]$, the first and second derivatives vanish identically and there is no rearrangement term in the HFB or in the QRPA matrix. This is the case of the \textit{volume} pairing, while \textit{mixed} and \textit{surface} parametrizations give rise to a non zero rearrangement term. 
 Its impact on the centroids ranges from 0.04 - 0.2 MeV. In particular, the energy weighted sum rules are very sensitive to the correct handling of the rearrangement term. This term should thus be properly taken into account for a calculation to be considered fully self consistent.
 As an example we report in Fig. \ref{fig:rearr} the strength functions for $^{112}$Sn with and without the rearrangement term. The energy weighted sum rule in the full calculation including the rearrangement term is 99.2\%. If we neglect the rearrangement in both of the HFB and QRPA it is 137\%, while neglecting the rearrangement term only in the QRPA calculation leads to EWSR exhausted at 116\%. The effect on the strength function is very patent for this nucleus where, in the fully self consistent calculation, the peak at 16 MeV is higher than the one at 17 MeV, while in the calculations without rearrangement term the opposite happens; as a result these last calculations tends to return higher centroids respect to the fully self consistent case. 
 
\subsection{QRPA numerical precision}
For a good comparison with the experimental data it is useful to constrain the numerical uncertainty originating from the calculations.   
Three main parameters influence the quality of the QRPA calculations: the mesh size, the box size and the quasiparticle energy cutoff. At first we analyze how the centroids of $^{132}$Sn are affected by the mesh size; Tab. \ref{tab:mesh} shows that this parameter has a very small impact on the results and we decided to stick to 0.05 fm mesh since it is often used in literature.

  \begin{table}[ht!]
     \begin{center}
        \begin{tabular}{|c|c|c|c|}
        \hline   
\multicolumn{4}{|c|}{ $^{132}$Sn, ~~ SkM* + \textit{volume} }\\ 
\hline
 mesh (fm)    &~~ $m_1/m_0$ ~~ &~~  $ (m_3/m_{-1})^{1/2}$ ~~  &~~$(m_1/m_{-1})^{1/2}$ ~~\\
\hline
0.025    &     15.37  &       15.30  &        15.56 \\ 
\hline
0.05     &     15.36  &       15.30  &        15.55 \\
\hline
0.1      &     15.36  &       15.29  &        15.54 \\
\hline
0.2      &     15.36  &       15.29  &        15.54 \\
\hline 
        \end{tabular} 
     \end{center}
\caption{Centroids (MeV) calculated in the energy range 10.5 - 20.5 MeV as a function of the mesh size for $^{132}$Sn. The box radius is fixed at 20 fm, the cutoff is 60 MeV.}
\label{tab:mesh}
\end{table}

The next step is to vary the box size. The bigger the box the better one expects the continuum to be approximated. However, this is at the cost of increasing the number of states of the QRPA matrix and making the diagonalization slow. 

 As a good compromise we decided to use a 20 fm box. Going to higher box sizes would increase the precision of the results by about 0.03 Mev but the two quasiparticle space would increase by 1.6 times (25 fm box) rendering the calculations $\approx$ 4 times slower (since the diagonalization of the QRPA matrix scales as the third power of the matrix size).

  \begin{table}[ht!]
     \begin{center}
        \begin{tabular}{|c|c|c|}
        \hline   
\multicolumn{3}{|c|}{ $^{132}$Sn, ~~ SkM* + \textit{volume} }\\ 
\hline
 box size (fm)    &~~ $m_1/m_0$ ~~  & ~~ 2qp ~~\\
\hline
    18   &  15.29    &  2746  \\
\hline
    20   &  15.36   &  3460  \\
\hline
    25   &  15.33   &  5640  \\
\hline 
    30   &  15.33   &  8320  \\
\hline    
        \end{tabular} 
     \end{center}
\caption{Centroids (MeV) calculated in the 10.5 - 20.5 MeV range as a function of the box size for $^{132}$Sn and the size of the two quasiparticle space (the QRPA matrix size is two times this quantity). The mesh is 0.05 fm and the cutoff is 60 MeV.}
\label{tab:box-size}
\end{table}
At this point we address the size of the quasiparticle space as a function of the cutoff.
A space too small can lead to not fully self consistent results. All the quasiparticle excitations obtained in the HFB calculation are also employed in the QRPA calculations (this is necessary for full self consistency).
No prescription is used to take into account the continuum other than using a box size big enough to obtain stable results.
 As can be seen in Tab. \ref{tab:2qpspace}, stability is reached in the cutoff range 150~-~200 MeV.
  \begin{table}[ht!]
     \begin{center}
        \begin{tabular}{|c|c|c|c|c|}
        \hline   
\multicolumn{4}{|c|}{ $^{132}$Sn, ~~ SkM* + \textit{volume} }\\ 
\hline
cutoff        &~~   $m_1/m_0$   ~~       &~~  $(m_1/m_{-1})^{1/2}$ ~~  &~~$(m_3/m_1)^{1/2}$ ~~ \\
  \hline
60    &   15.36  &      15.30   &     15.55    \\
\hline
100   &   15.30  &      15.24   &     15.48    \\
\hline
150   &   15.26  &      15.20   &     15.43    \\
\hline
200   &   15.25  &      15.19   &     15.43    \\ 
\hline
        \end{tabular} 
     \end{center}
\caption{The centroid of the ISGMR (MeV), as a function of the energy cutoff (MeV) of the two quasiparticle space. The moments are calculated in the energy interval 10.5 - 20.5 MeV. The box is 20 fm wide with a 0.05 fm mesh.}
\label{tab:2qpspace}
\end{table}

In summary, according to the preceding results, we decide to employ in the rest of the article a 20 fm box with 0.05~fm mesh and a 200 MeV cutoff. These parameters allow us to estimate the numerical uncertainty of our calculations to be $<$~0.1 MeV.

 The value of the pairing gap, in the case of DDDI pairing interaction is a function of the cutoff. In order to obtain gap values corresponding to the experimental five point formula we adjusted the values of the strength $V_0$ as detailed in Appendix \ref{App:A}. 
 The resulting QRPA two quasiparticle space usually contains about $\approx$~6000 states leading to matrices of the order of $\approx$ 12000$\times$ 12000 entries. 
The spurious states which can appear at low energy ($<$~1~MeV) have been proven to be very well separated from the physical modes \cite{mFam} and we can thus ignore them.
  \begin{table}[ht!]
     \begin{center}
        \begin{tabular}{|c|c|c|c|}
\hline
\multicolumn{4}{|c|}{ $^{174}$Sn, ~~ SkM* + \textit{volume} } \\
\hline
                       &~~    $m_1/m_0$~~          &  $(m_1/m_{-1})^{1/2}$ & $(m_3/m_1)^{1/2}$  \\ 
\hline 
Our                                              &  12.925     &   12.849 &       13.185  \\
\hline
Terasaki \textit{et al.}  \cite{Terasaki1}       &  12.924     &   12.848 &       13.184 \\
\hline
        \end{tabular} 
     \end{center}
\caption{Moment ratios (MeV) evaluated in the range 10.5 - 20.5 MeV obtained with our calculations and those from \cite{Terasaki1} cutoff (iii), for $^{174}$Sn SkM* with \textit{volume} pairing. The box is 20 fm with a 0.05 fm mesh, the 2qp cutoff is 200 MeV.} \label{tab:comparisonTera}
\end{table}
In order to check the robustness of our results we compare them with the fully self consistent calculations by Terasaki \textit{et al.} \cite{Terasaki1}. The two QRPA approaches are completely different 
(FAM \cite{mFam} and canonical basis expansion \cite{Terasaki1}), and they only share the underlying HFB. Despite this fact, the difference in the centroids is about 0.001 MeV as can be seen in Table \ref{tab:comparisonTera}.

\subsection{Experimental settings}
Experiments on giant monopole resonances have been performed for more than 30 years.
The number and accuracy of the data obtained in recent years is increasing, and for this reason we will limit our analysis to the most recent experiments. 
Two main groups have been recently working on ISGMR, namely the team at Texas A\&M (TAMU) at College Station (USA), and the team of the Research Center for Nuclear Physics (RCNP) at Osaka University (Japan). Many detailed analysis of giant resonances have been carried out but, there are still some discrepancies which have to be addressed. 
In the following, we will review the major findings of both groups,
however, since the quantities to be compared are functions of the moments of Eq. (\ref{eq:moment}) and different experiments use non equivalent systems to evaluate them, they will be addressed separately.
In particular the energy range of evaluation of the moments changes from experiment to experiment, and this can affect the centroids. 
\section{Lead}\label{sec:lead}
   \begin{table}[ht!]
     \begin{center}
        \begin{tabular}{|c|c|c|c|}
        \hline   
\multicolumn{4}{|c|}{ $^{208}$Pb  }\\
\hline
  $m_1/m_0$  (MeV)      &    EWSR \%     & method & reference \\
\hline 
 14.17 $\pm$ 0.28       &                & slice analysis    & TAMU  \cite{ZrPRL1999}  \\
\hline
13.5  $\pm$ 0.2         &   76 $\pm$ 5   & Breit-Wigner      & RCNP \cite{SmZrPb} \cite{osaka208Pb}\\
\hline
 13.96 $\pm$ 0.20       &   99 $\pm$ 15  & slice analysis    & TAMU  \cite{Sm2004}    \\
\hline  
        \end{tabular} 
     \end{center}
\caption{ISGMR experimental centroids $m_1/m_0$ for $^{208}$Pb.} 
\end{table}
$^{208}$Pb is the paradigmatic nucleus on which the compressibility modulus has been studied for a long time \cite{Old-exp-Young}. For a summary of the latest experiments one can refer to \cite{Sm2004}.
%
%
There is a general agreement that the experimental strength of $^{208}$Pb is rather symmetric around the centroid located at 13.5 - 14 MeV with a width of 2.5 - 3~MeV. We will take as a reference the latest results found in \cite{Sm2004} for $^{208}$Pb since they almost completely exhaust the energy weighted sum rule (EWSR). The moments used for the centroid are calculated in the full energy range of the experiment, in this case being 10 - 55 MeV.
  \begin{table}[t!]
     \begin{center}
        \begin{tabular}{|c|c|c|c|c|c|}
        \hline   
\multicolumn{3}{|c|}{  Pb } &
\multicolumn{3}{|c|}{ SLy5 }\\
        \hline 
     N  &     Z   & ~~~~ Exp. ~~~~ &~~ \textit{volume}  ~~&~~ \textit{mixed}~~&~~ \textit{surface}~~\\ 
\hline 
  122   &     82  &  13.98 &         \BF{13.91}     &       13.90       &     13.84           \\ 
  124   &     82  &  13.94 &         \BF{13.86}     &       13.85       &     13.84           \\
  126   &     82  &~~~~  13.96$^{+0.2}_{-0.2}$ &   \BF{13.85}     &       13.84       &  13.85   \\
        \hline
        \hline
\multicolumn{3}{|c|}{      } &
\multicolumn{3}{|c|}{ SkM* }\\
        \hline 
     N  &     Z   & ~~~~ Exp. ~~~~ &~~ \textit{volume} ~~&~~ \textit{mixed}~~&~~  \textit{surface}~~\\ 
\hline 
122     &   82  & 13.98   &   13.51  &   13.49   &   13.45  \\
124     &   82  & 13.94   &   13.45  &   13.44   &   13.43  \\ 
126     &   82  &~~~~ 13.96$^{+0.2}_{-0.2}$   &   13.43  &   13.43   &   13.43  \\
\hline
        \hline   
 \multicolumn{3}{|c|}{        } &
 \multicolumn{3}{|c|}{ Skxs20 }\\
        \hline 
     N  &     Z   & ~~~~ Exp. ~~~~ &~~ \textit{volume} ~~&~~ \textit{mixed} ~~&~~ \textit{surface}~~\\ 
\hline 
122     &   82  &     13.98                   &  13.22  &   13.19  &    13.12   \\
124     &   82  &     13.94                   &  13.17  &   13.16  &    13.07   \\ 
126     &   82  &~~~~ 13.96$^{+0.2}_{-0.2}$   &  13.12  &   13.12  &    13.08*  \\
\hline
        \end{tabular} 
     \end{center}
\caption{ Pb centroids $m_1/m_0$ (MeV) calculated with the SLy5, SkM* and Skxs20 interactions with \textit{volume}, \textit{mixed} and \textit{surface} pairing interactions. The energy range for the calculations of the moments is 10 - 55 MeV. The experimental results are from Ref. \cite{Fujiwara, Sm2004}.
(*) Although $^{208}$Pb is doubly magic, the neutron pairing gap resulting from the HFB calculations for Skxs20 and \textit{surface} pairing is 0.62 MeV.}
\label{Tab:Pb}
\end{table}
 No energy range is explicitly stated for the experimental values of the centroids of $^{204-206}$Pb observed at RCNP \cite{Fujiwara}, however, since the strength of Pb isotopes is narrowly peaked, the centroids are expected to have a limited dependence on the energy range. For this reason we stick, for homogeneity, to the 10 - 55 MeV energy range.  
Our results about Pb isotopes confirm that the HFB+QRPA framework reproduces very well 
the experimental findings when using a SLy5 interaction ($K_{NM}\approx$ 230 MeV). The centroids in the case of the SkM* interaction ($K_{NM}\approx$ 217 MeV) are underestimated by about $0.5$ MeV. The Skxs20 interaction ($K_{NM} \approx$ 202 MeV) returns values about 0.8 MeV lower than the experiment. The difference between the 
\textit{volume, mixed} and \textit{surface} pairing interactions is $\leq$ 0.1 MeV. 
These results confirm the generally accepted picture, according to which Pb can be reproduced with a good approximation by using an interaction with incompressibility modulus in nuclear matter of about 230 MeV.
\section{Tin}\label{sec:Tin}
The interactions which reproduce well $^{208}$Pb (e.g. SLy5) tend to overestimate the centroids of Sn isotopes \cite{Pieka07}.
This unexpected softness of Sn isotopes introduced new doubts on the almost settled question of the incompressibility of atomic nuclei. 

 \subsection{Comparison with RCNP data}

 \begin{figure}[ht!]
 \centering
 \includegraphics[trim = 7mm 4mm 5mm 5mm, width=8cm]{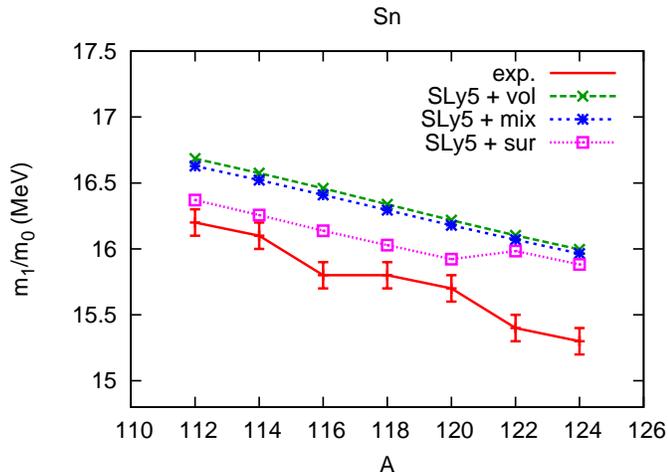}
 \caption{(Color online) Sn experimental centroids $m_1/m_0$ (MeV) from \cite{Sn-exp-osaka} compared to the calculations with SLy5 interaction with \textit{volume}, \textit{mixed} and \textit{surface} pairing. The energy range for the calculation of the moments is 10.5 - 20.5 MeV for both of experiments and calculations.}
  \label{fig:sn-sly}
 \end{figure}
 \begin{figure}[ht!]
 \centering
 \includegraphics[trim = 7mm 4mm 5mm 5mm, width=8cm]{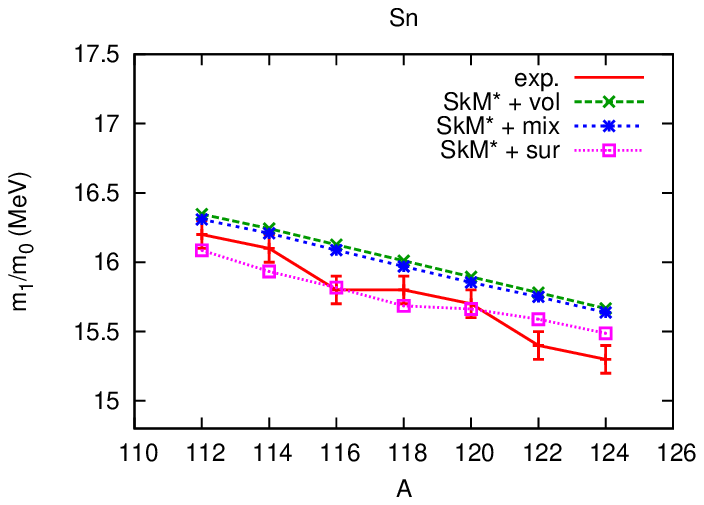}
 \caption{(Color online) As in Fig. \ref{fig:sn-sly} but using the SkM* Skyrme interaction in the ph channel.}
  \label{fig:sn-skm}
 \end{figure}
 \begin{figure}[ht!]
 \centering
 \includegraphics[trim = 7mm 4mm 5mm 5mm, width=8cm]{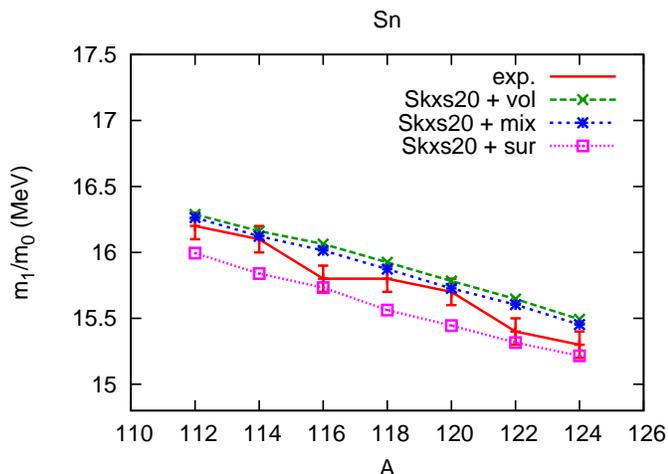}
 \caption{(Color online) As in Fig. \ref{fig:sn-sly} but using the Skxs20 Skyrme interaction in the ph channel.}
  \label{fig:sn-skp}
 \end{figure}
Seven Sn isotopes have been observed at RCNP \cite{Sn-exp-osaka}. 
The measured centroids tend to diminish as a function of the number of nucleons, however the trend is marked by some staggering phenomena. For example, the centroid of $^{116}$Sn is 0.3 MeV lower than the centroid of $^{114}$Sn, but $^{116}$Sn and $^{118}$Sn show the same centroid. It is possible to divide the observed centroids in two groups: a stiff one containing $^{112-114-118-120}$Sn, and a soft one containing $^{116-122-124}$Sn. 

SLy5 Skyrme interaction shows higher centroids than the experiments by about 0.5 MeV (see Fig. \ref{fig:sn-sly}). This difference can be partially cured (by about 0.3 MeV) with use of \textit{surface} pairing interactions. Experimentally the decrease of the centroid from  $^{120}$Sn to $^{122}$Sn is rather steep, while the calculations with SLy5 and \textit{surface} pairing, on the other hand, show the opposite behavior with $^{122}$Sn being stiffer than $^{120}$Sn. As a result no pairing interaction can allow a SLy5 calculation to obtain results in good agreement with the experiments.

 SkM* with \textit{volume/mixed} and  \textit{surface} interaction are in rather good agreement for many of the isotopes under consideration, while the agreement decreases for $^{122-124}$Sn (Fig. \ref{fig:sn-skm}). 
The correspondence with the experiments is particularly good for $^{116}$Sn (SkM* and \textit{surface} pairing produces the closest description of this isotope among the ones used in this work) and $^{120}$Sn.

Skxs20 and \textit{surface} interaction reproduces very well $^{116-122-124}$Sn while the same interaction with \textit{volume} parametrization gives the best approximation for $^{112-114-118-120}$Sn (Fig. \ref{fig:sn-skp}). In general, despite the fact that the centroids obtained with the SkM* are in good agreement with the experiments, the Skxs20 interaction gives the best approximation of the experimental data. The calculations for Sn isotopes seem less sensitive to the value of $K_{NM}$ than for other nuclei (e.g. lead), which implies that a smaller gap between the experiments and theoretical data is required to obtain a reliable value of $K_{NM}$. 
%
%
%
%
\subsection{Comparison with TAMU data} 
The agreement on the experimental results of Sn isotopes is not complete. In Tab. \ref{tab:Sn-exp}
 we compare the data from RCNP \cite{Sn-exp-osaka} and TAMU \cite{Sn-exp1}~\cite{Sm2004}:
  \begin{table}[ht!]
     \begin{center}
        \begin{tabular}{|c|c|c|c|}
\hline
                 &   RCNP                         &          TAMU        &  diff. (MeV)    \\ 
\hline 
$^{112}$Sn       &  16.2$^{+0.1}_{-0.1}$  \cite{Sn-exp-osaka}    & 15.43$^{+0.1}_{-0.1}$  \cite{Sn-exp1}  & 0.77  \\ 
\hline
$^{116}$Sn       &  15.8$^{+0.1}_{-0.1}$   \cite{Sn-exp-osaka}   & 15.85$^{+0.2}_{-0.2}$    \cite{Sm2004} & -0.05 \\ 
\hline
$^{124}$Sn       &  15.3$^{+0.1}_{-0.1}$   \cite{Sn-exp-osaka}   & 14.50$^{+0.14}_{-0.14}$  \cite{Sn-exp1}& 0.8 \\
\hline
        \end{tabular} 
     \end{center}
\caption{The Sn experimental centroids $m_1/m_0$ (MeV) have been calculated in the energy range 10.5 - 20.5 MeV for  \cite{Sn-exp-osaka} while the range is 10 - 55 MeV for \cite{Sm2004}\cite{Sn-exp1}.} \label{tab:Sn-exp}
\end{table}

Two of RCNP's centroids ($^{112}$Sn and $^{124}$Sn) are higher than the corresponding ones observed at TAMU, while very good agreement is reached for $^{116}$Sn. A source of uncertainty when comparing the centroids of the two different experimental groups is that they do not use the same prescription for extracting the moments. The Osaka group (for Sn and Cd isotopes) restricts the energy interval of Eq. (\ref{eq:moment}) to 10.5 - 20.5 MeV.  
One of the reasons for this limited energy interval is an observed spurious strength at high energies, see for example Ref. \cite{Sn-exp-2010} and \cite{exp-cadmium2}.
This spurious plateau is attributed to continuum effects mimicking the L=0 angular momentum component of the total excitation strength.
To our knowledge there is no numerical estimate indicating the range where the spurious continuum strength 
significantly affects the experimental results. If a tail of the spurious strength extends at low energies (affecting the interval where the moments are calculated) this would lead to an artificial increase of the calculated centroids. Conversely, the upper limit of the energy range of the moments (20.5 MeV) might exclude some of the strength at high energy, thus lowering the ``real'' values. This latter hypothesis seems the most reasonable situation since we calculated the theoretical centroids in different energy ranges. When the calculation of the momenta was passing from the energy range 10.5 - 20.5 MeV to 10 - 55 MeV the centroids increased by about 0.2 MeV. If this was the case also for the experimental values obtained in \cite{Sn-exp-osaka} the discrepancy with respect to TAMU's results would increase. 
  \begin{table}[ht!]
     \begin{center}
        \begin{tabular}{|c|c|c|c|}
\hline
                     &    Exp. \cite{Sn-exp1}   & Skxs20 + \textit{surface}  & diff. (MeV)   \\ 
\hline 
$^{112}$Sn           &    15.43$^{+0.1}_{-0.1}$        &  16.21    &    0.78    \\ 
\hline
$^{116}$Sn           &    15.85$^{+0.2}_{-0.2}$        &  15.88    &    0.03     \\ 
\hline
$^{124}$Sn           &    14.50$^{+0.14}_{-0.14}$      &  15.37    &    0.87     \\
\hline
        \end{tabular} 
     \end{center}
\caption{Sn centroids $m_1/m_0$ (MeV) of \cite{Sn-exp1} compared with the Skxs20 + \textit{surface} interaction, in the last column we show the difference between the theoretical results and the experimental ones. The moments are evaluated in the range 10 - 55 MeV.} \label{tab:Sn-Tamu}
\end{table}

The centroids observed in \cite{Sn-exp1} are very low and they seem to correspond to parametrizations much softer compared to the ones used to reproduce most of other observed nuclei. Among the Skyrme forces we use, the only reasonable comparison can be carried out with the Skxs20 + \textit{surface} pairing. Nonetheless the theory overestimates the experiments by about 0.8 - 0.9 MeV (Tab. \ref{tab:Sn-Tamu}). 
Notice that the centroids calculated with Skxs20 + \textit{surface} shown in Tab. \ref{tab:Sn-Tamu} are $\approx$~0.2~MeV higher than the  corresponding values displayed in Fig.~\ref{fig:sn-skp} because of the different ranges where they have been evaluated (10 - 55 MeV for the former 10.5 - 20.5 MeV for the latter).
\section{Zirconium}\label{sec:Zirconium}      
  \begin{table}[ht!]
     \begin{center}
        \begin{tabular}{|c|c|c|c|c|c|}
        \hline   
\multicolumn{3}{|c|}{ Zr } & 
\multicolumn{3}{|c|}{ SLy5 }\\
        \hline 
     N  &     Z   & ~~~~ Exp. ~~~~ &~~ \textit{volume}  ~~&~~ \textit{mixed}~~&~~ \textit{surface}~~\\ 
\hline 
   50   &     40  &  17.88$^{ +0.13}_{ -0.11}$ &\BF{18.04}  &    18.04       &    18.04      \\ 
   52   &     40  &  18.23$^{ +0.15}_{ -0.13}$ &\BF{17.84}  &    17.83       &    17.77      \\
   54   &     40  &  16.16$^{ +0.12}_{ -0.11}$ &    17.62   &    17.62       &    17.59      \\
        \hline       
        \hline   
\multicolumn{3}{|c|}{      } & 
\multicolumn{3}{|c|}{ SkM* }\\
        \hline 
     N  &     Z   & ~~~~ Exp. ~~~~ &~~ \textit{volume}  ~~&~~ \textit{mixed}~~&~~ \textit{surface}~~\\ 
\hline 
   50   &     40  & 17.88$^{ +0.13}_{ -0.11}$  &    17.64        &    17.63          &     17.62    \\ 
   52   &     40  & 18.23$^{ +0.15}_{ -0.13}$  &    17.43        &    17.41          &     17.41    \\
   54   &     40  & 16.16$^{ +0.12}_{ -0.11}$  &    17.18        &    17.17          &     17.18    \\
        \hline       
        \hline   
\multicolumn{3}{|c|}{        } & 
\multicolumn{3}{|c|}{ Skxs20 }\\
        \hline 
     N  &     Z   & ~~~~ Exp. ~~~~ &~~ \textit{volume}  ~~&~~ \textit{mixed}~~&~~ \textit{surface}~~\\ 
\hline 
   50   &     40  & 17.88$^{ +0.13}_{ -0.11}$  &  17.53          &   17.49     &    17.34       \\ 
   52   &     40  & 18.23$^{ +0.15}_{ -0.13}$  &  17.36          &   17.28     &    17.12       \\
   54   &     40  & 16.16$^{ +0.12}_{ -0.11}$  &  17.19          &   17.14     &\BF{17.02}       \\
\hline        \end{tabular} 
     \end{center}
\caption{Zr centroids $m_1/m_0$ (MeV) calculated with the SLy5, SkM* and Skxs20 interactions and \textit{volume}, \textit{mixed} and \textit{surface} pairing interactions. The energy range for the calculations of the moments is 10 - 55 MeV. The experimental results are from Ref.~\cite{ZrShlomo}.}
\label{Tab:Zr}
\end{table}
The experimental centroid of $^{90}$Zr can be reproduced with a good approximation with the same theoretical framework which successfully reproduces $^{208}$Pb (see Tab. \ref{Tab:Zr}).
The experimental strength function, at variance in respect to $^{208}$Pb is asymmetrical.
Two papers published by the Texas A\&M University group in 1999 \cite{ZrPRL1999} and 2004 \cite{ZrPRC2004} agree with the fact that the tail of the resonance extends beyond 25 MeV.
This is due to the fact that the giant monopole resonance for $^{90}$Zr \cite{ZrShlomo} shows two peaks (one approximately at $16 - 17$ MeV and another one at about $24 - 25$ MeV) while in our calculations there is essentially only one main peak. 

We report in Tab. \ref{Tab:Zr} a comparison between our calculations and the experimental values of \cite{ZrPRC2004, ZrShlomo}.
According to these calculations the SLy5 and SkM* forces reproduce well the experimental value of $m_1/m_0$ for $^{90}$Zr, while the Skxs20 underestimates it by about 0.3 MeV. For this isotope the choice of the pairing interaction plays a minor role in determining the centroid of the ISGMR.

Recently new experimental data on $^{92-94}$Zr isotopes have been presented in \cite{ZrShlomo}. The findings for $^{92}$Zr are particularly different respect to $^{90}$Zr, since this former nucleus is very stiff. It is, therefore, best approximated by the SLy5 interaction, and the difference from the experimental and theoretical values of the centroid is about 0.4 MeV.

On the other hand, $^{94}$Zr is surprisingly softer than $^{92}$Zr by about 2 MeV, and even the softest interaction adopted for this nucleus (Skxs20 with \textit{surface}) overestimates the centroid by about 0.8 MeV.
The calculated centroids, independent of the interaction used in the ph or in the pp channel, show a rather smooth behavior as a function of A, and cannot reproduce the observed experimental staggering among adjacent isotopes. This situation is similar to the observed inversion of the centroids of Ca isotopes \cite{Ca40-exp, Ca-exp}. $^{40}$Ca, in fact, is significantly softer than $^{48}$Ca. These nuclei have been recently studied within the HF+RPA approximation by Anders \textit{et al.} in \cite{Ca40-48} by checking which of the Skyrme parameters can account for the inversion in the centroids, however no interaction was able to reproduce it.

In the present calculation the nuclei are treated as spherical. Strictly speaking, however, all the nuclei display some degree of deformation which can affect the response function of the giant monopole resonance. In the case of the nuclei under investigation the deformation parameter is rather small ($\beta <0.2$) \cite{deformation} thus we expect deformation to play a minor role. However, an investigation making use of deformed QRPA might help in explaining the reason of the observed staggering between $^{92}$Zr and $^{94}$Zr.

In the present work we remain within the QRPA method which is a harmonic
approximation. As such, by construction it does not take into account the
anharmonicities which might affect the excitation spectrum. A quantitative
estimate of the anharmonicities is difficult because it involves the
analysis of the coupling between different modes, as for example shown in
ref. \cite{ColoPLB}. Blaizot \textit{et al.} in \cite{Blaizot591} estimated
it with the help of constrained HFB calculations. It was shown that the
importance of anharmonicities is higher for lighter nuclei, such as Ca,
while it is expected to be negligible for heavy nuclei such as Pb. This
effect might also be partially responsible for the gap between the stiffness
of $^{92}$Zr and $^{94}$Zr and it suggests the need for further
investigations with models beyond the QRPA.
\section{Samarium}\label{sec:Samarium} 
  \begin{table}[t!]
     \begin{center}
        \begin{tabular}{|c|c|c|c|c|c|c|}
        \hline   
\multicolumn{7}{|c|}{ Sm } \\
        \hline 
     N  &     Z   & ~~~ Exp. ~~~ & ph int.& ~~ \textit{volume}  ~~&~~ \textit{mixed}~~&~~ \textit{surface}~~\\
\hline
   82   &     62  &  15.40$^{+0.3}_{-0.3}$& SLy5   & 15.69    &  15.58     &  15.65*       \\ 
\hline 
   82   &     62  &  15.40$^{+0.3}_{-0.3}$& SkM*   & \BF{15.27}    &  15.24     &  15.17*       \\
\hline 
   82   &     62  &  15.40$^{+0.3}_{-0.3}$& Skxs20 & 15.02    &  14.99     &  14.76*       \\
\hline
 \end{tabular} 
     \end{center}
 \caption{$^{144}$Sm $m_1/m_0$ ratios (MeV), the moments are calculated in the energy interval 10 - 55 MeV. (*) \textit{Surface} pairing calculations return a non zero neutron pairing gap in contrast with the fact that 82 is a magic number.} 
\label{Tab:Sm}
\end{table}
The RCNP and TAMU groups have been investigating different Sm isotopes.
There is good agreement about the experimental properties of the ISGMR of $^{144}$Sm \cite{ZrPRL1999, Sm144-154, Sm2004}; this resonance displays a single peak whose centroid is expected to be around 15.3~-~15.4 MeV with a width of about 4 MeV. According to \cite{Sm2004} the observed strength is almost zero beyond 20 MeV. The calculations with a SkM* and SLy5 reproduce the experimental values within the experimental errors (see Tab. \ref{Tab:Sm}). Even though the Skxs20 interaction tends to underestimate the centroids, the average distance between the lower bound experimental values and the theoretical results obtained with \textit{volume} and \textit{mixed} pairing is limited to about 0.1 MeV. A very interesting article \cite{Sm2003} reported a number of Sm isotopes, however as the neutron number 
increases they become more and more deformed. This leads to a mixing of the monopole and quadrupole resonance and for this reason a careful theoretical investigation should make use of deformed QRPA methods like those reported in \cite{axialTera,Yoshida2010,CriFam}.   
Since $^{144}$Sm is neutron magic one expects that the neutron pairing gap of a HFB calculation should be zero (we tuned the pairing interaction strength $V_0$ to reproduce the experimental value of the proton pairing gap). In the case of the \textit{surface} pairing interaction, for all the three forces, the neutron pairing gap is ``unnaturally'' different from zero and approximately 1.2 MeV. For this reason a broad margin of error has to be taken into account for  \textit{surface} pairing for this chemical element.
\section{Molybdenum}\label{sec:Molybdenum}    
  \begin{table}[ht!]
     \begin{center}
        \begin{tabular}{|c|c|c|c|c|c|}
        \hline   
\multicolumn{3}{|c|}{   Mo   } & 
\multicolumn{3}{|c|}{ SLy5 }\\
        \hline 
     N  &     Z   & ~~~~ Exp. ~~~~ &~~ \textit{volume}  ~~&~~ \textit{mixed}~~&~~ \textit{surface}~~\\ 
\hline 
   50   &     42  & 19.62$^{ +0.28}_{ -0.19}$  & \BF{18.01}      &   17.99     &   17.96       \\ 
   54   &     42  & 16.95$^{ +0.12}_{ -0.10}$  &  17.58      &   17.56     &   17.53       \\
        \hline       
        \hline   
\multicolumn{3}{|c|}{      } & 
\multicolumn{3}{|c|}{ SkM* }\\
        \hline 
     N  &     Z   & ~~~~ Exp. ~~~~ &~~ \textit{volume}  ~~&~~ \textit{mixed}~~&~~ \textit{surface}~~\\ 
\hline 
   50   &     42  & 19.62$^{ +0.28}_{ -0.19}$  &  17.59      &  17.57      &   17.53       \\ 
   54   &     42  & 16.95$^{ +0.12}_{ -0.10}$  &  17.24      &  17.22      &   17.19       \\
        \hline       
        \hline   
\multicolumn{3}{|c|}{      } & 
\multicolumn{3}{|c|}{ Skxs20 }\\
        \hline 
     N  &     Z   & ~~~~ Exp. ~~~~ &~~ \textit{volume}  ~~&~~ \textit{mixed}~~&~~ \textit{surface}~~\\ 
\hline 
   50   &     42  & 19.62$^{ +0.28}_{ -0.19}$  &     17.47    &  17.42      &  17.25        \\ 
   54   &     42  & 16.95$^{ +0.12}_{ -0.10}$  &     17.13    &  17.07      &  \BF{16.95}        \\
\hline        \end{tabular} 
     \end{center}
\caption{Mo centroids $m_1/m_0$ (MeV) calculated with the SLy5, SkM* and Skxs20 interactions and \textit{volume}, \textit{mixed} and \textit{surface} pairing interactions. The energy range for the calculations of the moments is 10 - 55 MeV. The experimental results are from Ref.~\cite{ZrShlomo}.}
\label{Tab:Mo}
\end{table}
We report in Tab. \ref{Tab:Mo} the results of the calculations for Molybdenum obtained with SLy5, SkM* and Skxs20 Skyrme interactions, and the experimental results of \cite{ZrShlomo}.
These experiments with Mo isotopes are very recent and add new information on nuclear incompressibility.
In particular $^{92}$Mo represents a challenging problem because of its very high stiffness. 
All the interactions we use cannot explain the experimental strength of $^{92}$Mo. Even the stiffest interaction we adopted, the SLy5, fails at reproducing its centroid by about 1.5 MeV by defect. This difference is particularly challenging since the problem up to now was to explain too soft nuclei compared with $^{208}$Pb, while $^{92}$Mo has a completely opposite behavior. An additional experimental confirmation would be very useful for this nucleus. 
On the other hand the experiments for $^{96}$Mo are very well reproduced with Skxs20 + \textit{surface}  interaction and also the SkM* interaction returns centroids only slightly higher than the experiments.
\section{Cadmium}\label{sec:Cadmium}
\subsection{Comparison with RCNP data}
 \begin{figure}[ht!]
 \centering
 \includegraphics[trim = 7mm 4mm 5mm 5mm, width=8cm]{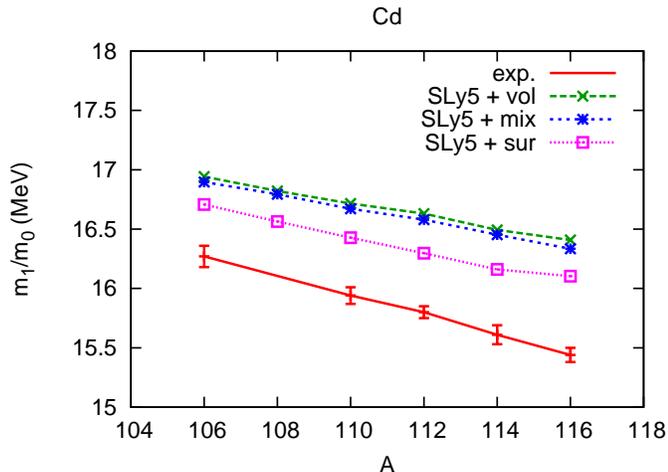}
 \caption{(Color online) Cadmium centroids $m_1/m_0$ calculated with SLy5 interaction and the experimental centroids from Ref. \cite{exp-cadmium2}. The energy range used for the evaluation is 10.5 - 20.5 MeV. The pairing interactions are of the \textit{volume}, \textit{mixed} and \textit{surface} type. }
  \label{fig:cd-sly5}
 \end{figure}
 \begin{figure}[ht!]
 \centering
 \includegraphics[trim = 7mm 4mm 5mm 5mm, width=8cm]{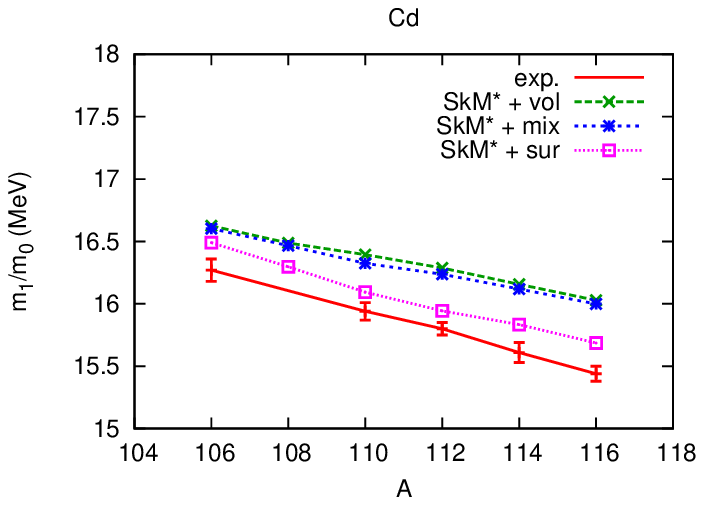}
\caption{(Color online) As in Fig. \ref{fig:cd-sly5} but using the SkM* interaction in the ph channel.}
  \label{fig:cd-skm}
 \end{figure}
 \begin{figure}[ht!]
 \centering
 \includegraphics[trim = 7mm 4mm 5mm 5mm,  width=8cm]{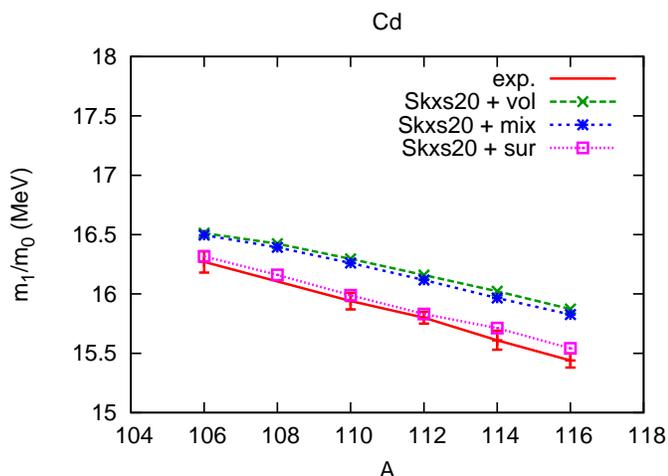}
\caption{(Color online) As in Fig. \ref{fig:cd-sly5} but using the Skxs20 interaction in the ph channel.}
  \label{fig:cd-skx}
 \end{figure}

Cd isotopes have shown similar properties as Sn isotopes in 
recent experiments at RCNP \cite{exp-cadmium2, exp-cadmium3}. In our calculations with SLy5 (and \textit{volume/mixed} pairing) interaction, the ISGMR centroids are in general higher than these experiments by about 0.7 MeV (Fig. \ref{fig:cd-sly5}). The \textit{surface} pairing lowers the centroids by about 0.2 - 0.3 MeV but this is not enough for an accurate reproduction of the experiments. Also, the SkM* (\textit{volume/mixed}) calculations, as shown in Fig. \ref{fig:cd-skm}, are stiffer than the experiments by 0.4 - 0.6 MeV. \textit{Surface} pairing partially reduces the problem to about 0.2 - 0.3 MeV.
Both of the experiments and the theoretical values of the centroids decrease by increasing the number of nucleons.
The Skxs20  (plus \textit{volume/mixed} pairing) returns values slightly softer than the SkM* thus diminishing the gap with the experiments. The most accurate reproduction of the experiments is obtained when applying the \textit{surface} pairing. With this choice of the parameters the difference between experiments and theory is lower than 0.05 MeV for $^{106-110-112}$Cd. 
 The quality decreases slightly for $^{114-116}$Cd but in any case the agreement is about 0.1 MeV (Fig. \ref{fig:cd-skx}).
  \begin{table}[ht!]
     \begin{center}
        \begin{tabular}{|c|c|c|c|}
\hline
             &  RCNP       &        TAMU         & diff (MeV)   \\ 
\hline 
$^{110}$Cd   &  15.94$^{+0.07}_{-0.07}$ \cite{exp-cadmium2} &    15.12$^{+0.11}_{-0.11}$  \cite{Cd-exp-TAMU} & 0.82   \\ 
\hline
$^{116}$Cd   &  15.44$^{+0.06}_{-0.06}$ \cite{exp-cadmium2} &    14.50$^{+0.16}_{-0.16}$   \cite{Cd-exp-TAMU}& 0.94   \\
\hline
        \end{tabular} 
     \end{center}
\caption{Comparison of the different experimental centroids $m_1/m_0$ (MeV) for Cd isotopes. The moments for the RCNP results are evaluated in the range 10.5 - 20.5 MeV, while those of TAMU are in 10 - 55 MeV. } 
\label{Tab:Cd-RCNP-TAMU}
\end{table}

\subsection{Comparison with TAMU data}
The experimental uncertainty that was present for Sn isotopes is also present for cadmium. The TAMU group reports centroid values \cite{Cd-exp-TAMU} significantly lower than the ones obtained at RCNP \cite{exp-cadmium3}. For this reason, the only meaningful comparison with these experiments is with the softest parametrization possible (Skxs20 + \textit{surface}). Even though this returns the best approximation among the ones we tested, it highly overestimates the experiments as can be seen in Tab. \ref{tab:Cd-TAMU}. Notice that also in this case the centroids of Tab. \ref{tab:Cd-TAMU} are different in respect to the ones of Fig. \ref{fig:cd-skx}. This is because of the dependence of the energy range where they have been calculated (10.5~-~20.5 MeV when comparing with RCNP results and 10~-~55 MeV when comparing with TAMU results).  

  \begin{table}[ht!]
     \begin{center}
        \begin{tabular}{|c|c|c|c|}
\hline
                     &    Exp. \cite{Cd-exp-TAMU}   & Skxs20 + \textit{surface}  & diff (MeV)   \\ 
\hline 
$^{110}$Cd           &    15.12$^{+0.11}_{-0.11}$      &  16.20     &   1.08         \\ 
\hline
$^{116}$Cd           &    14.50$^{+0.16}_{-0.16}$      &  15.76     &   1.26          \\
\hline 
        \end{tabular} 
     \end{center}
\caption{The experimental \cite{Cd-exp-TAMU} and calculated $m_1/m_0$ (MeV) for Cd isotopes. The energy range for the calculation is 10 - 55 MeV. All values are in MeV. } \label{tab:Cd-TAMU}
\end{table}
%
%
%
%
%
%
%
%
\section{Differences with respect to HF+BCS}\label{sec:confrontoColo}
It is useful to discuss the differences between our work and the one by Cao \textit{et al}. \cite{colo+cadmium} since the approach to the problem is similar, but the conclusions drawn contain differences. The main theoretical difference between the two approaches stems from the usage of a HFB or a HF + BCS calculation as a basis for the QRPA. 

In general the results presented in Ref. \cite{colo+cadmium} for \textit{volume/mixed} pairing are in good agreement with the ones we find for SLy5 and SkM*, the difference being only 0.1~-~0.2 MeV. 

The major discrepancy regarding the pairing interactions is related to \textit{surface} pairing.
In all of our calculations the \textit{surface} pairing returns softer nuclei with respect to \textit{volume/mixed} pairing, at variance in the work by Cao \textit{et al.} \textit{surface} pairing returns stiffer nuclei by 0.2 - 0.3 MeV respect to \textit{volume/mixed} pairing. 
In order to understand the reason of this difference it is useful to refer to previous articles, for example Li \textit{et al.} \cite{colo+hfb+qrpa} performed a HFB + QRPA calculation of Sn isotopes. Their results are qualitatively similar to ours (although in Ref. \cite{colo+hfb+qrpa} there is no two-body spin-orbit term and no rearrangement term due to the pairing interaction). Li \textit{et al.} also find that \textit{surface} pairing softens superfluid nuclei more than \textit{mixed/volume} pairing. The same qualitative behavior is reported  in the constrained HFB calculation presented by Khan \textit{et al.} in \cite{Khan-Sagawa}. %
Since the works which report softer \textit{surface} are based on HFB calculations while \cite{colo+cadmium} is based on HF + BCS we think that this should be the reason for the discrepancy. The HFB theory is a generalization of the HF + BCS framework and we expect that the former should lead to a more realistic description of superfluid nuclei.
Because of the different qualitative behavior, the agreement between ours and Ref. \cite{colo+cadmium} centroids, for \textit{surface} pairing, is rather poor leading to differences of about 0.4 MeV.

 Another important difference is to be searched in the mean field used in the calculations.
  In Ref. \cite{colo+cadmium} the SkP interaction was chosen to check whether soft parametrizations ($K_{NM} \approx$ 200 MeV) could give a good reproduction of the centroids. This Skyrme interaction, however, was shown to lead to false ground states in Ref. \cite{ref-err-skp}. This state appears stable even when using high precision for the convergence of the mean field, while the ``real'' ground state presents nonphysical oscillations.
  The mean field properties obtained with the SkP interaction are, thus, not stable and we decided to use as representative of the soft parametrizations the Skxs20 interaction which was produced to replace the SkP among the soft interaction in Ref. \cite{ref-skx}.

All these facts lead to a different overall picture, since in our case the soft Skxs20 interaction with \textit{surface} pairing returns the best results among the forces we investigated, while in Ref. \cite{colo+cadmium} the soft parametrization was the worst among the chosen Skyrme forces and \textit{surface} pairing gave in general poorer results compared to \textit{volume/mixed}
pairing.

\section{Isovector Pairing interactions}\label{sec:isovector-pairing}

In this section we will address the effect of different kinds of pairing interactions which display more complicated density dependence respect to Eq. (\ref{eq:pairing-int}).

All of the interactions so far presented have an isoscalar approach to pairing, and they do not take into account explicitly the asymmetry between neutrons and protons. 
In order to better reproduce the pairing gap in nuclei, recently, isovector pairing interactions have been proposed \cite{MSH, YS} and returned good agreement with the experimental data \cite{isovBert}.

The isovector pairing interaction, denoted by MSH \cite{MSH} is parametrized as follows:
\begin{equation}
  \begin{split}
 v^{MSH}_{pair}({\bf r},{\bf r'})  = & 
     V_0 \! \left[1  - (1 - \delta)\eta_s \left( \frac{\rho}{\rho_0} \right)^{\alpha_s} \!  -\delta\eta_n\left(\frac{\rho}{\rho_0}\right)^{\alpha_n}\right] \\ 
     & \times \delta({ \bf r}-{\bf r'}),
    \end{split}
\end{equation}
In its original version \cite{MSH} $\rho=\rho_n+\rho_p$, $\delta= (\rho_n-\rho_p)/\rho$, $V_0=-448$ MeV fm$^{3}$, $\eta_s=0.598$, $\alpha_s=0.551$, $\eta_n=0.947$ and $\alpha_n=0.554$ (the cutoff being 60 MeV, $\rho_0=0.16$ fm$^{-3}$).
%
%
%
%
 \begin{figure}[ht!]
  \centering
  \includegraphics[trim = 7mm 4mm 5mm 5mm, width=8cm]{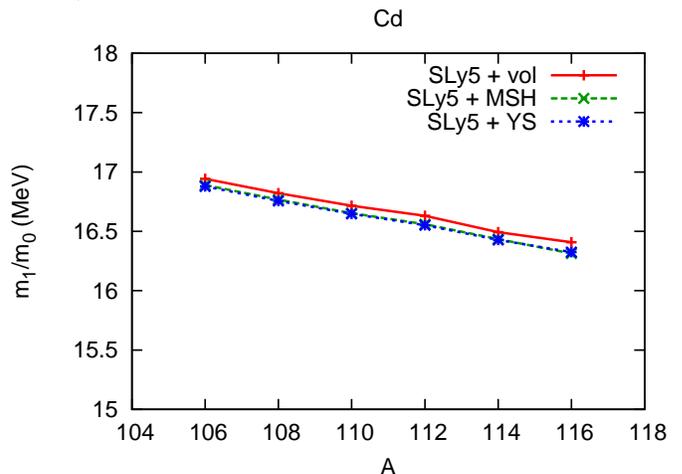}
  \caption{(Color online) Cd centroids $m_1/m_0$ (MeV), calculated with SLy5 interaction and \textit{volume}, MSH and YS pairing. The energy range for the moments is 10.5 - 20.5 MeV.}
   \label{fig:Cd-iso}
  \end{figure}

 \begin{figure}[ht!]
  \centering
  \includegraphics[trim = 7mm 4mm 5mm 5mm, width=8cm]{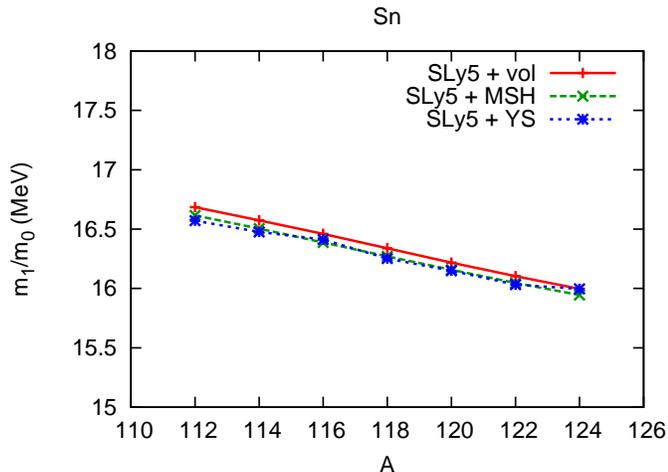}
  \caption{(Color online) As in Fig. \ref{fig:Cd-iso} but for Sn isotopes.}
   \label{fig:Sn-iso}
  \end{figure}
A different pairing parametrization, denoted by YS, is provided in \cite{YS}:
\begin{equation}
 \begin{split}
 v^{YS}_{pair}({ \bf r},{ \bf r}') = & 
           V_0
                                 \left[1-\left(\eta_0+\eta_1\tau_3\delta\right) \frac{\rho}{\rho_0}
                                -\eta_2\left( \delta \frac{\rho}{\rho_0} \right)^{2}   \right]  \\
   & \times  \delta({ \bf r}-{\bf r}'),
  \end{split}
\end{equation}
where the parameter set proposed in \cite{YS} (we use \cite{Khan-Sagawa} as a summary for the values of the interactions) is $V_0=-344$ MeV fm$^{3}$, $\eta_0=0.5$, $\eta_1=0.2$ and $\eta_2=2.5$ (with an energy cutoff of 50 MeV), $\tau_3= 1$ for neutrons and -1 for protons. 
In Fig. \ref{fig:Cd-iso} and Fig. \ref{fig:Sn-iso} we show the Cd and Sn isotopes calculated with SLy5 interaction in the particle hole channel and \textit{volume}, MSH and YS pairing.
Since the cutoff we employ is different in respect to the original parametrizations, we adjust the parameter $V_0$ according to the nuclei we want to study. When using the MSH parametrization we choose $V_0$=-317.5 MeV f$m^{3}$ for Cd isotopes and -335.85 MeV fm$^{3}$ for Sn isotopes. In the case of YS pairing we set $v_0=$ -253.1 MeV fm$^{3}$ for Cd and -268.6 MeV fm$^{3}$ for Sn.
 The pairing fields returned by these parametrizations resemble the ones with \textit{volume/mixed} pairing and it is thus no surprise that the centroids of the ISGMR are in good agreement with these interactions. On the overall there is no strong effect on the centroids due to the isovector dependence of the pairing interaction. Isovector pairing does not provide a good solution to the problem related with the nuclear incompressibility.
\section{Extreme pairing interactions}\label{sec:Pairing}
So far we had a further confirmation that pairing correlations generally tend to soften the ISGMR.   
Given the flexibility of our calculations, we want to test pairing correlations to their limit
(even going beyond realistic kinds of pairing) to check if it exists a pairing parametrization that can decrease even further the differences between the experiments and the theory.
At first we will focus to test if the pairing interaction always softens nuclei. In the following we change the value of the strength of the pairing interaction $V_0$. The resulting pairing gap is not, in general, close to the experimental values; so these calculations are to be thought only as hints of the effect of pairing on the nuclear structure. 
We used the Skyrme SkM* with \textit{volume} pairing for $^{120}$Sn.
When we pass from 0.65 MeV pairing gap to the realistic 1.39 MeV there is a slight increase of the stiffness. The stiffness increases further when passing to a $\Delta_n$= 1.99 MeV, and there is a very strong enhancement of the stiffness for the (totally unrealistic) $\Delta_n$= 5.49 MeV.  
This calculation suggests that the effect of pairing within a Skyrme functional framework does not 
necessarily soften a nucleus.

  \begin{table}[ht!]
     \begin{center}
        \begin{tabular}{|c|c|c|}
\hline
\multicolumn{3}{|c|}{ $^{120}$Sn, SkM*+ \textit{volume}  }\\
\hline
 $V_0$ (MeV fm$^{3}$) & $\Delta_n$ (MeV) & $m_1/m_0$ (MeV) \\
\hline
-100.0    &  0.65    &  15.88   \\
-132.8    &  1.39    &  15.89   \\
-150.0    &  1.99    &  15.97   \\
-200.0    &  5.49    &  16.49   \\  
\hline
        \end{tabular} 
     \end{center}
\caption{Centroids as a function of the strength of the pairing interaction ($V_0$); the moments have been calculated in the energy interval 10.5 - 20.5 MeV. } 
\label{tab:gap-strength}
\end{table}

\begin{figure}[ht!]
 \centering
\includegraphics[trim = 7mm 0mm 4mm 0mm, width=8cm]{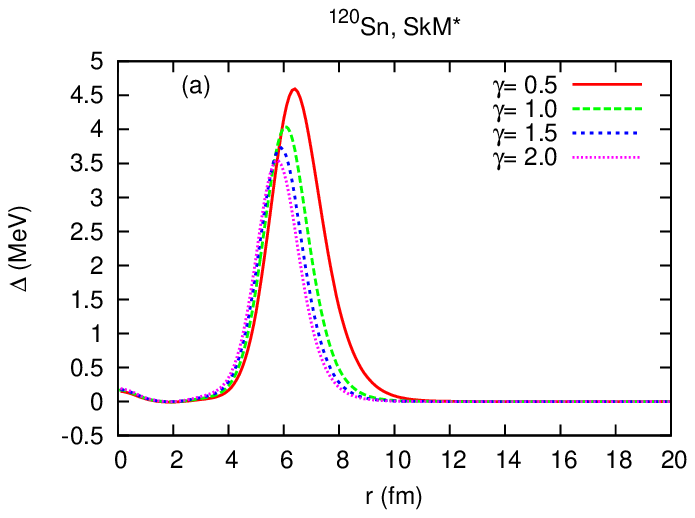}
\includegraphics[trim = 7mm 0mm 4mm 0mm, width=8cm]{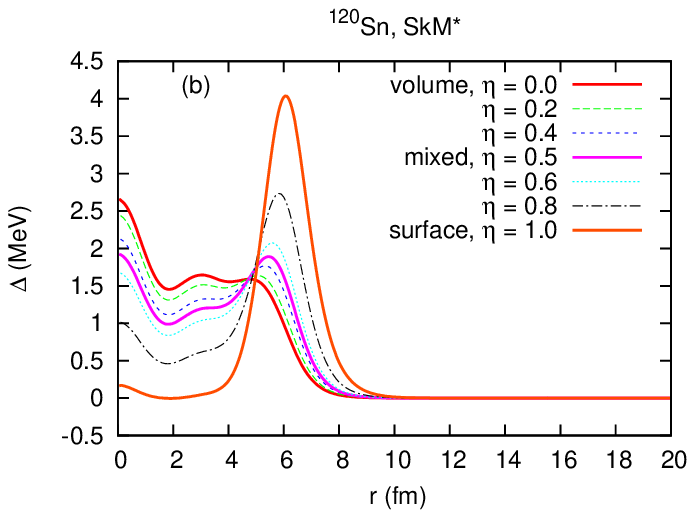}
 \caption{(Color online) (a) \textit{Surface} pairing fields ($\eta$=1) obtained by changing the exponent $\gamma$ of the density dependence and adjusting $V_0$ to obtain the same neutron pairing gap. (b) Pairing fields obtained by changing the value of $\eta$, and tuning $V_0$ in order to obtain the same neutron pairing gap.}
  \label{fig:Pairing-comparison}
 \end{figure}

We decided to test the pairing interactions as a function of the parameter $\eta$. The most reasonable values of this parameter are $\eta\in[0,1]$ since at the extremes we have the \textit{volume} and \textit{surface} pairing and for $\eta=0.5$ \textit{mixed} pairing is obtained. In Fig. \ref{fig:Pairing-comparison} (b) are shown the different pairing fields passing from \textit{volume} to \textit{surface} pairing.  
A change of  $\eta$ requires to adjust $V_0$ in order to keep the pairing gap consistent as reported in Tab. \ref{tab:eta}.
Our results show that the softest parametrization of the pairing field as a function of $\eta$  
is obtained with \textit{surface} pairing. According to Tab. \ref{tab:eta} values of $\eta$ between 0 and 0.6 return very similar centroids while the biggest changes are obtained only very close to $\eta=1$, that is \textit{surface} pairing. One could be tempted to extrapolate even further this trend, however values of $\eta>1$ are more difficult to interpret since they would mean a situation of anti-pairing inside the nucleus and a very strong pairing condition on the nuclear surface.

  \begin{table}[ht!]
     \begin{center}
        \begin{tabular}{|c|c|c|c|}
\hline
\multicolumn{4}{|c|}{ $^{120}$Sn, SkM*  }\\
\hline
 $V_0$ (MeV fm$^{3}$) & $\eta$ & $\gamma$  & $m_1/m_0$ (MeV)  \\
\hline
   -132.8 &   0.0  & 1.0     &  15.89  \\            
   -157.5 &   0.2  & 1.0     &  15.89  \\   
   -191.8 &   0.4  & 1.0     &  15.87  \\   
   -214.2 &   0.5  & 1.0     &  15.85  \\                               
   -241.5 &   0.6  & 1.0     &  15.84  \\   
   -313.0 &   0.8  & 1.0     &  15.74  \\    
   -404.5 &   1.0  & 1.0     &  15.66    \\                                        
\hline
        \end{tabular} 
     \end{center}
\caption{Centroids $m_1/m_0$ (calculated in the 10.5 - 20.5 MeV energy range) as a function of the parameter $\eta$. The pairing strength $V_0$ and the value of $\gamma$ are also shown.}
 \label{tab:eta}
\end{table}

Since \textit{surface} pairing shows the most important effects on the centroid of the ISGMR, this seems to suggest that low density regions can be important in softening nuclei.
We decided to modify the value of the exponent in the density dependence of the surface interaction in order to change the position of the peak of the pairing field.

In the calculations shown in Table \ref{tab:pairing} we restrict to \textit{surface}  pairing ($\eta$=1). Also in this case the strength $V_0$ was adjusted to keep the pairing gap constant for $^{120}$Sn.
\begin{table}[ht!]
     \begin{center}
        \begin{tabular}{|c|c|c|c|c|}
        \hline
\multicolumn{4}{|c|}{ $^{120}$Sn, SkM*  }\\
        \hline   
                           $V_0$ (MeV fm$^{3}$) &  $\eta$    &   $\gamma$  & $m_1/m_0$ (MeV)     \\
\hline
                           -608.0      &  1.0       &     0.5           &   15.72    \\
\hline     
                           -404.5      &  1.0       &     1.0           &   15.66    \\
\hline      
                           -331.5      &  1.0       &     1.5           &   15.60    \\
\hline
                           -293.3      &  1.0       &     2.0           &   15.58    \\
\hline
        \end{tabular} 
     \end{center}
\caption{Centroids for \textit{surface} pairing ($\eta$ = 1) interaction as a function of different values of the exponent $\gamma$, and the resulting values of $V_0$ needed to keep the value of the pairing gap constant. The moments are evaluated in the 10.5 - 20.5 MeV energy range.}
\label{tab:pairing} 
\end{table}
The resulting $m_1/m_0$ ratios are essentially similar to the centroids of the standard \textit{surface} pairing interaction. When $\gamma=0.5$ the nucleus is slightly stiffer than the case with $\gamma=1$. As one can see from Fig. \ref{fig:Pairing-comparison} when $\gamma=2.0$ the pairing field is about 1 fm at the right of the standard $\gamma $= 1 pairing and the centroid is about 0.1 MeV softer.
These results suggest that the region in between the surface and the center of the nucleus is the most sensitive for the determination of the compression properties. However, the general result is that the pairing associated to Eq. (\ref{eq:pairing-int}), when constrained to reasonable values of the pairing gap, does not lead to dramatic changes of the incompressibility. 
%
%
%
\section{Conclusion}\label{sec:Conclusion}
\begin{table}[ht!]
     \begin{center}
        \begin{tabular}{|c|c|c|c|}
        \hline   
 nucleus                 &  ph              & pp                                &  diff.            \\
\hline 
$^{204-206-208}$Pb       &  SLy5            &    all                            &       $<$ 0.1     \\
\hline
\hline   
$^{144}$Sm               &  SkM*            &    \textit{volume}                &         - 0.1     \\
\hline 
\hline     
$^{90}$Zr                &  SLy5            &    all                            &         + 0.2     \\
$^{92}$Zr                &  SLy5            &    \textit{volume}                &         - 0.4     \\
$^{94}$Zr                &  Skxs20          &    \textit{surface}               &         + 0.8     \\
\hline
\hline
$^{92}$Mo                &  SLy5            &    \textit{volume}                &         - 1.6     \\
$^{96}$Mo                &  Skxs20          &    \textit{surface}               &         + 0.0     \\
\hline
\hline
$^{112-114-118-120}$Sn \cite{Sn-exp-osaka}  &  Skxs20   &    \textit{mixed}     &        $<$ 0.1    \\
$^{122-124}$Sn       \cite{Sn-exp-osaka}    &  Skxs20   &    \textit{surface}   &        $<$ 0.1    \\
$^{116}$Sn           \cite{Sn-exp-osaka}    &  SkM*     &    \textit{surface}   &        $<$ 0.1    \\

$^{112-124}$Sn   \cite{Sn-exp1}             &  Skxs20   &    \textit{surface}   & $\approx$  0.8    \\
$^{116}$Sn   \cite{Sn-exp1}                 &  Skxs20   &    \textit{surface}   &         + 0.2     \\
\hline
\hline
$^{106-110-112-114-116}$Cd \cite{exp-cadmium3}&  Skxs20 &    \textit{surface}   &        $<$ 0.1    \\
$^{110-116}$Cd \cite{Cd-exp-TAMU}           &  Skxs20   &    \textit{surface}   &          + 1    \\ 
\hline 
        \end{tabular} 
     \end{center}
\caption{We summarize the interactions which best reproduce the experimental data and we report the average difference between the calculations and experimental findings in MeV (positive value meaning that the calculations overestimate the experiments); a ``$<$'' symbol indicates that the isotopes are within the given range independent on the sign of the difference. A ``all'' in the pp column means that the pairing interaction plays a marginal role in the determination of the centroids. In case of ``remarkable" difference among the experimental results we display both of them and the best theoretical approximation associated.}
\label{tab:summary} 
\end{table}

We have presented the first fully self consistent HFB+QRPA+DDDI calculation that takes into account the rearrangement due to the density dependence of the pairing interaction, and we compared it with the state of the art experiments on isoscalar giant monopole resonances.

When comparing the centroids of different Skyrme interactions there is a rather strong correlation with the associated value of $K_{NM}$. Increasing by $\approx$ 15 MeV the value of $K_{NM}$ in turn increases the value of the centroids by 0.2~-~0.4 MeV. Conversely, this means that the precision on the calculations is essential, since neglecting effects which introduce an uncertainty on the centroids $\geq$ 0.2 MeV can suggest choices of Skyrme parametrizations differing in the value of $K_{NM}$ by up to $\approx$ 15 MeV.

For this reason we put particular attention on minimizing the sources of uncertainty when dealing with superfluid nuclei.

In the case of magic nuclei pairing plays essentially no role in determining the properties of the system,
while in the majority of the superfluid nuclei we studied \textit{surface} pairing gives the best approximation to the experimental values.
The effect of \textit{surface} pairing is generally to ``soften'' the ISGMR with respect to \textit{volume} and \textit{mixed} parametrizations by about 0.2 - 0.3 MeV. 

We pushed the DDDI interactions to their limits, but it seems that this type of pairing alone is not capable to account for a unique parametrization for all the studied nuclei.
This is valid also for the isovector dependent pairing interactions which we have introduced for the first time in a QRPA calculation. The centroids obtained with MSH and YS pairing are, in fact, compatible with \textit{volume/mixed} pairing. 

While the theoretical centroids vary rather smoothly with the number of nucleons, some very recent experiments  display strong variations between neighboring nuclei. These new features, if confirmed with further experimental tests, would require important modifications of the present models used to reproduce the ISGMR. It is particularly interesting that the two isobars $^{92}$Mo and $^{92}$Zr are the stiffest nuclei so far observed and this opens the question whether other A = 92 nuclei might display an unusual stiffness.

Although about 20\% of the nuclei under investigation are well explained with the SLy5 interaction, and 10\% with the SkM* interaction, it is remarkable that, for the majority of the nuclei under investigation (see Tab. \ref{tab:summary}), the ISGMR is better reproduced with the soft interaction Skxs20 ($K_{NM}\approx$ 202 MeV) in contrast with the generally accepted value for $K_{NM}\approx 230$ MeV. The uncertainty deriving from the different experimentally determined centroids (which can be up to $\approx$ 1 MeV) leads to even softer general pictures. All these results show that there is need for reconsideration of the generally accepted value of the incompressibility of infinite systems.

\appendix

\section{Pairing Strength} \label{App:A}
The pairing gap obtained with a contact interaction depends on the size of the quasiparticle space and this latter is a function of the used cutoff, in our case 200 MeV. For this reason we select one isotope for each chemical element and we modify $V_0$ in order to reproduce the experimental pairing gap given by the five point formula \cite{Audi2003} according to Tab. \ref{Tab:sly5-V0}. \\

%
\begin{table}[H]
     \begin{center}
        \begin{tabular}{|c|c|c|c|c|c|}                                                         
\hline
  \multicolumn{6}{|c|}{   Skxs20   }\\   
\hline
  &   N&  Z  &   volume  $V_0$ & mixed $V_0$  & surface $V_0$ \\      
\hline
Cd&  64&   48&         -112.0  &     -189.0   &  -380.0      \\
Mo&  54&   42&         -126.5  &     -209.8   &  -380.0     \\
Pb& 122&   82&         -128.5  &     -219.0   &  -453.5      \\
Sm&  82&   62&         -128.0  &     -222.8   &  -463.5     \\
Sn&  70&   50&         -113.8  &     -193.0   &  -394.0      \\
Zr&  54&   40&         -121.2  &     -201.0   &  -358.0      \\
\hline                                                          
  \multicolumn{6}{|c|}{   SkM*   }\\                                                            
\hline    
         &   N&  Z  &   volume  $V_0$ & mixed $V_0$  & surface $V_0$ \\      
\hline
Cd&  64 &  48 &      -130.0  &    -229.7&     -387.0       \\      
Mo&  54 &  42 &      -148.5  &    -233.5&    -391.0        \\
Pb& 122 &  82 &      -143.5  &    -237.0&     -467.5       \\     
Sm&  82 &  62 &      -161.5  &    -268.5&     -461.0       \\     
Sn&  70 &  50 &      -132.8  &    -214.2&     -404.5       \\   
Zr&  54 &  40 &      -141.5  &    -222.0&     -368.0       \\      
\hline
  \multicolumn{6}{|c|}{   SLy5   }\\                      
\hline
  &   N&  Z  &   volume  $V_0$ & mixed $V_0$  & surface $V_0$ \\      
\hline
Cd&  64&   48&     -152.8  &      -238.2 &    -419.8   \\
Mo&  54&   42&     -156.0  &      -240.8 &    -401.8   \\
Pb& 122&   82&     -156.5  &      -253.0 &    -493.0   \\
Sm&  82&   62&     -173.3  &      -285.4 &    -490.0   \\
Sn&  70&   50&     -159.2  &      -248.7 &    -441.0   \\             
Zr&  54&   40&     -148.0  &      -225.3 &    -372.3   \\             
\hline
        \end{tabular} 
     \end{center}
\caption{Pairing strength $V_0$ (MeV fm$^{3}$) for the different Skyrme and pairing interactions and the isotope used for the calibration.}
   \label{Tab:sly5-V0}
\end{table}

\section{Skyrme Properties}\label{App:B}
  \begin{table}[ht!]
     \begin{center}
         \begin{tabular}{|c|c|c|c|c|c|c|c|c|c|c|}
\hline
Skyrme  & $\rho_0$ &  $E_0$ & $K_{NM}$   &    $J$   &     $L$  & $K_{sym}$ &  $m^*/m$ \\ 
\hline
SLy5   &  0.161 &  -15.99 &  229.92  &  32.01 &  48.15 &  -112.76 &    0.70  \\
SkM*   &  0.160 &  -15.77 &  216.61  &  30.03 &  45.78 &  -155.94 &    0.79  \\
Skxs20 &  0.162 &  -15.81 &  201.95  &  35.50 &  67.06 &  -122.31  &   0.96  \\                                      
\hline
        \end{tabular} 
     \end{center}
\caption{Nuclear matter properties associated to the different Skyrme interactions.}
 \label{Tab:NM}
\end{table}

In Tab. \ref{Tab:NM} we show the main nuclear matter properties associated to the Skyrme interactions (a detailed explanation of the quantities of Tab. \ref{Tab:NM} can be found in Ref. \cite{stevenson}). 
In particular we provide the saturation density $\rho_0$, the binding energy at saturation $E_0$, the incompressibility of nuclear matter ($K_{NM}$), the first expansion terms of the symmetry energy ($J$, $L$ and $K_{sym}$) and the effective mass $m^*/m$.
The symmetry energy is defined as: 
\begin{equation}
 \mathcal{S}= \frac{1}{8} \left. \frac{\partial^{2} (\mathcal{E}/\rho)}{\partial {y}^2} \right|_{\rho,y=1/2}
\end{equation}
where $y=\rho_p/\rho$ is the proton fraction and $\mathcal{E}$ is the energy density. This quantity can be expanded as:
\begin{equation}
 \mathcal{S}= J+Lx+\frac{1}{2}K_{sym}x^{2}+O(x^{3})
\end{equation}
being $x=(\rho-\rho_0)/3\rho_0$.\\

\begin{figure}[ht!]
  \centering
  \includegraphics[trim = 7mm 2mm 5mm 0mm, width=8cm]{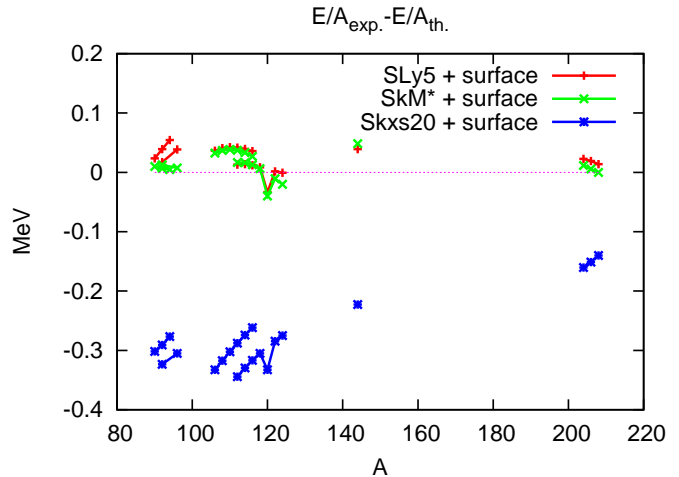}
  \caption{(Color online) Difference between the experimental \cite{Audi2003} and the theoretical binding energies (isotopes of the same element are connected by a curve).}
   \label{fig:binding}
  \end{figure}

\begin{figure}[ht!]
  \centering
  \includegraphics[trim = 7mm 2mm 5mm 0mm, width=8cm]{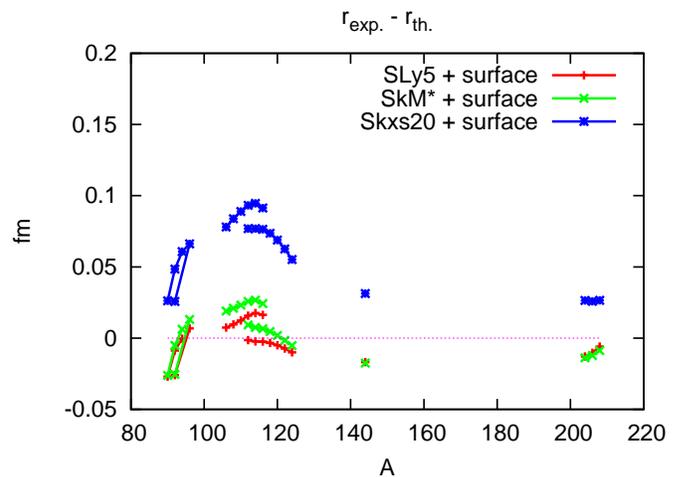}
  \caption{(Color online) Difference between the experimental \cite{radii} and the theoretical charge radii (isotopes of the same element are connected by a curve).}
   \label{fig:radii}
  \end{figure}

We report the experimental binding energies and charge radii of the nuclei under consideration in Tab. \ref{Tab:radii}.
\begin{table}[ht!]
     \begin{center}
        \begin{tabular}{|c|c|c|c|c|}                                                         
\hline
 N&  Z  &   Nucleus  & charge radius (fm)  & E/A (MeV) \\      
\hline 
50  &  40 &  Zr     &      4.27         &     8.71          \\
52  &  40 &  Zr     &      4.31         &     8.69          \\
54  &  40 &  Zr     &      4.33         &     8.67          \\
\hline                                            
50  &  42 &  Mo     &      4.32         &     8.66          \\                
54  &  42 &  Mo     &      4.38         &     8.65          \\
 \hline                                           
58  &  48 &  Cd      &     4.53         &     8.54          \\
60  &  48 &  Cd      &     4.55         &     8.55          \\
62  &  48 &  Cd      &     4.57         &     8.55          \\
64  &  48 &  Cd      &     4.59         &     8.54          \\
66  &  48 &  Cd      &     4.61         &     8.53          \\
68  &  48 &  Cd      &     4.63         &     8.51          \\
  \hline                                          
62  &  50 &  Sn    &       4.59       &       8.51          \\
64  &  50 &  Sn    &       4.61       &       8.52          \\
66  &  50 &  Sn    &       4.63       &       8.52          \\
68  &  50 &  Sn    &       4.64       &       8.52          \\
70  &  50 &  Sn    &       4.65       &       8.47          \\
72  &  50 &  Sn    &       4.67       &       8.49          \\
74  &  50 &  Sn    &       4.68       &       8.47          \\
 \hline                                           
82  &  62 &  Sm    &       4.94       &       8.30          \\ 
 \hline                                           
122 &  82 &  Pb     &      5.48       &       7.88          \\ 
124 &  82 &  Pb     &      5.49       &       7.88          \\  
126 &  82 &  Pb     &      5.50       &       7.87          \\   
\hline

        \end{tabular} 
     \end{center}
\caption{Experimental charge radii (fm) and binding energy (KeV) from \cite{radii} and \cite{Audi2003}.}
   \label{Tab:radii}
\end{table}
 In Fig.~\ref{fig:binding} we show the difference between the experimental and calculated binding energies for the three Skyrme interactions. In Fig.~\ref{fig:radii} we show the difference between the experimental and calculated charge radii. Since the calculated energies and radii for \textit{volume}, \textit{mixed} and \textit{surface} pairing interaction are similar to each other we report only the latter case.
The SLy5 and SkM* preform better results respect to the Skxs20 on both of the binding energies and charge radii. The Skxs20, in fact, overestimates the binding energy and returns smaller nuclei with respect to the experiments. Nonetheless the number of low $K_{NM}$ Skyrme parametrizations available is rather limited and the Skxs20 is the only one that can pass most of the macroscopic constraints of ref. \cite{stevenson}. 
\\ 
%

\section*{Acknowledgments}\label{sec:ack}
P.A. would like to thank T. Nakatsukasa for a careful reading of the manuscript and for the useful advices; A. Pastore for his suggestions about the SkP interaction and S. Shlomo for the discussions about full self consistency. This work is supported by the US-DOE grants: DE-SC004971 and DE-FG02-08ER41533.

\end{document}